# Lattice dynamical calculations of infinite layer iron oxides $SrFeO_2$ and $CaFeO_2$

R. Mittal[1], S. L. Chaplot[2] and Y. Su[1]

[1]Jülich Centre for Neutron Science, Institut für Festkörperforschung, Forschungszentrum Jülich, Outstation at FRM-II, Lichtenbergstrasse 1, D-85747 Garching, Germany

[2]Solid State Physics Division, Bhabha Atomic Research Centre, Trombay, Mumbai- 400085, India

## Abstract

We report extensive lattice dynamical calculations of the newly discovered infinite-layer iron oxides $SrFeO_2$ and $CaFeO_2$. For $SrFeO_2$, the parameters of the interatomic potential have been determined to reproduce the zone-centre phonon frequencies reported using ab-initio calculations. Further we have extended the potential model for calculations of $CaFeO_2$. The potential parameters are found to be transferable between the two compounds, and are used to calculate the phonon spectra in the whole Brillouin zone and several thermodynamic properties for these compounds. The calculations show fair agreement with the available experimental data of structure, thermal expansion, and mean-squared amplitudes of the atoms.

Iron forms a large number of industrially useful oxides. In most oxides iron is found in tetrahedral or octahedral coordination. Until recently mineral gillespite $BaFeSi_4O_{10}$ was the only example [1] where iron has the square planar coordination of oxygen atoms, which is, however, interspersed by four-membered rings of $SiO_4$ tetrahedra. A family of $SrFeO_y$ and $CaFeO_y$ compounds (2.5 <y<3.0) has been known [2] with structures consisting of layers of $FeO_4$ tetrahedra and $FeO_6$ octahedra. Recently a new infinite-layer iron oxides $SrFeO_2$ has been synthesized [3,4]. The compound has an unusual square-planar coordination of oxygen atoms around iron. The tetrahedral structure (Fig. 1) of $SrFeO_2$ (space group P4/mmm) consists of layers of corner-sharing $FeO_4$ squares with strontium atoms in between. The structure closely resembles the sheet-like lattice geometry of superconducting copper oxides. There is no known structural instability as a function of temperature in $SrFeO_2$ and a similar compound $CaFeO_2$ [5].

$SrFeO_2$ exhibits exotic magnetic and electronic properties. Neutron diffraction studies show the presence of antiferromagnetic ordering with very high Neel temperature ($T_N$=473 K). The magnetic ordering temperature is very high in spite of the low-dimensional nature of the magnetic structure. Magnetic properties of $SrFeO_2$ have been examined [6,7] by density-functional-theory (DFT) band structure and total energy calculations. Ab-initio calculation [7] of the zone centre phonon modes has also been reported. It is of fundamental and practical importance to understand the stability of $SrFeO_2$ and $CaFeO_2$ as a function of temperature and pressure. We have carried out detailed lattice dynamical calculations to understand the phonon properties of these compounds.

Density functional theory calculations of the zone-center phonons using the local density approximation have been reported [7] for $SrFeO_2$. However we have used an atomistic approach based on empirical interatomic potentials. This approach enables calculations over a range of pressures and temperature using the same potential. Further, the potential parameters can be transferred for calculations in similar other compounds with suitable changes. The calculations were performed using an interatomic potential consisting of Coulombic and short-ranged terms. The interaction between two atoms $k$ and $k'$, separated by a distance $r$, is given by

$$V(r) = \{\frac{e^2}{4\pi\varepsilon_o}\}\{\frac{Z(k)Z(k')}{r}\} + a\exp\{\frac{-br}{R(k)+R(k')}\} - \frac{C}{r^6} - D\exp[-n(r-r_o)^2/(2r)] \quad (1)$$

where $r_{kk'}$ is the distance between the atoms $k$ and $k'$. The parameters of the interatomic potential are the effective charge $Z(K)$ and radius $R(k)$ of the atom type k. $1/(4\pi \varepsilon_o) = 9 \times 10^9$ Nm$^2$/Coul$^2$, a=1822 eV, b=12.364. The third term in eq.(1) is applied only between oxygen atoms. The radii parameters used in our calculations are R(Sr/Ca)= 2.07/2.03 Å, R(Fe)= 0.60 Å and R(O)=1.99 Å. Partial charges of Z(Sr/Ca)=2.00, Z(Fe)=1.0 and Z(O)= -1.50 are used in the calculations. The stretching potential given by the fourth term in Eq. (1) is applied only between the nearest Fe-O. The parameters of the stretching potential are D=4.25 eV, n=1.84 Å$^{-1}$, $r_o$ =1.966 Å.

The polarizibility of the oxygen atoms is introduced in the framework of the shell model [8,9] with the shell charge Y(Sr/Ca)=1.15, Y(O)= -1.2 and shell-core force constant K(Sr/Ca)=45 eV/Å$^2$, K(O)=100 eV/Å$^2$. Further the potential parameters are chosen such that they satisfy the conditions of static and dynamic equilibrium. The potential reproduces the structure of SrFeO$_2$ and CaFeO$_2$ as shown in Table I. The calculations are performed using the latest version of the program DISPR [12] developed at Trombay.

The unit cell of tetragonal SrFeO$_2$ (and CaFeO$_2$) has 4 atoms; thus there are 12 vibrational modes for each wave vector. Group theoretical symmetry analysis [13] was carried out to classify the phonon modes belonging to various representations. The phonon modes at the Γ point, and along the Δ, Λ and Σ directions decompose as follows:

Γ: 3A$_{2u}$ + B$_{2u}$ + 4E$_u$ (E$_u$ modes are doubly degenerate)
Δ: 4Δ$_1$ + 4Δ$_2$ + 4Δ$_3$
Λ: 4Λ$_1$+ 4Λ$_5$ (Λ$_5$ are doubly degenerate)
Σ: 4 Σ$_1$ + Σ$_2$ + 3Σ$_3$ + 4Σ$_4$

At the Γ point all the modes are infrared active. The shell model calculations and the available ab-initio calculations show the average deviation of only 2.2 % as seen in Table II. For CaFeO$_2$ no calculations are available. For CaFeO$_2$ we have only adjusted the radii parameter of Ca in the potential to reproduce the experimental structure. The calculated zone centre modes for CaFeO$_2$ are given in Table II. The phonon modes in CaFeO$_2$ are expected to have higher energies in comparison of SrFeO$_2$ due to the lower mass of Ca atoms and a smaller unit cell size of CaFeO$_2$. However, the B$_{2u}$ mode is found to have a lower frequency in CaFeO$_2$ in comparison of SrFeO$_2$. This mode involves out-of-phase motion of oxygen atoms along the crystallographic c-direction. The calculated phonon dispersion relation for SrFeO$_2$ along various high symmetry directions is shown in Fig. 2. Very large dispersion is obtained in several branches, especially in the highest energy $\Sigma_3$ branch that connects at the zone-boundary M-point (1/2,1/2,0) to the mode involving anti-phase rotations of FeO$_4$ tetrahedra.

The dynamical contributions to frequency distribution arising from the various species of atoms can be observed from their partial densities of states (Fig. 3), which are obtained by atomic projections of the one-phonon eigenvectors. The Sr and Ca atoms contribute largely up to 30 meV and 40 meV respectively. Due to the lighter mass of Ca (40.08 amu) in comparison to Sr (87.62 amu), the vibrations involving the Ca atoms are shifted toward higher energies. The contributions from Fe and O atoms are up to 50 and 80 meV, respectively. Above 60 meV the contributions are mainly due to Fe-O bond-stretching modes.

The calculated value of the bulk modulus for SrFeO$_2$ from our calculations is 130 GPa, which is in very good agreement with the reported ab-initio value [7] of 123 GPa, while the calculated value of bulk modulus for CaFeO$_2$ is 133 GPa. The experimental value of bulk modulus for SrFeO$_3$ is 147 GPa [14]. In SrFeO$_3$, Fe is six-coordinated. For the four-coordinated Fe compound, the bulk modulus is expected to be less than 147 GPa. So the calculated bulk modulus value of 130 GPa seems to be reasonable for SrFeO$_2$. The crystal structures for these compounds at different pressures are calculated by minimization of the free energy in our calculations. The calculated equation of state for SrFeO$_2$ and CaFeO$_2$ is shown in Fig. 4(a). It can be seen that CaFeO$_2$ is slightly more compressible than SrFeO$_2$.

The partial densities of states have been used for the calculation of the isotropic temperature factors for various atoms at different temperatures. Table I gives the comparison between the calculated and experimental values. The calculated temperature factors for Sr or Ca, Fe and O are somewhat similar in both the compounds.

Our calculated values of the Debye temperature are shown in Fig. 4(b), which shows significant temperature dependence. The Debye temperature estimated [5] from fitting of Debye Grüneisen equation to the temperature dependence of volume up to 300 K for SrFeO$_2$ and CaFeO$_2$ is 395 K and 733 K respectively. However, this involved the assumption of an equal Grüneisen parameter for all phonons.

The volume dependence of phonon energies has been used for the calculations of Grüneisen parameter. The Grüneisen parameters averaged over the whole Brillouin zone as a function of phonon energy (E) is calculated (Fig. 5(a)) by taking the contributions from calculated phonon frequencies at 405 wave vectors in the irreducible Brillouin zone. The mean Grüneisen parameters obtained for SrFeO$_2$ and CaFeO$_2$ are 1.12 and 1.13 respectively. The $\Gamma(E)$ values lie between 12.0 and 0.5. For CaFeO$_2$, $\Gamma(E)$ are very large, (>2) for phonons of energy below 10 meV. The calculated partial density of states (Fig. 2) shows that below 10 meV the contributions are due to Ca atoms.

The volume thermal expansion coefficient in the quasiharmonic approximation is given by $\alpha_V = \frac{1}{BV} \sum_i \Gamma_i C_{Vi}(T)$, where $\Gamma_i$ and $C_{Vi}$ are the mode Grüneisen parameter and the specific heat contributions of the phonons in state i(=**q**j, which refers to the $j^{th}$ phonon mode at wave vector **q**). The procedure of the calculation of thermal expansion is applicable when explicit anharmonicity of phonons is not very significant, and the thermal expansion arises mainly from the implicit anharmonicity, i.e., the change of phonon frequencies with volume. The higher order contribution to thermal expansion arising from variation of bulk modulus with volume [15, 16] is also included. The comparison between our calculations and experimental data of thermal expansion is shown in Fig. 5(b). The calculations for CaFeO$_2$ are in fair agreement with the available experimental data [5] while for SrFeO$_2$ there is a large difference between the calculations and experimental data. SrFeO$_2$ has antiferromagnetic ordering [3] with very high Neel temperature (T$_N$=473 K). Our calculation of thermal expansion does not include the contributions arising from the spin waves. For CaFeO$_2$ there is no magnetic transition reported in the literature.

In conclusion, our lattice dynamical model seems to be realistic to serve as a basis for an understanding of the phonon and thermodynamic properties of the infinite-layer iron oxides SrFeO$_2$ and CaFeO$_2$. The comparison between our calculated shell model and available ab-initio calculations of optic phonon frequencies is very good. The calculations of mean-squared amplitudes and thermal expansion are also in fair agreement with reported experimental data.

**Table I.** Comparison of the calculated structural parameters of $SrFeO_2$ and $CaFeO_2$ with the experimental data (Refs. [3] and [5]) at 293 K

|  | $SrFeO_2$ Expt. [3] | $SrFeO_2$ Expt [5] | $SrFeO_2$ Calc. | $CaFeO_2$ Expt. [5] | $CaFeO_2$ Calc. |
|---|---|---|---|---|---|
| a(Å) | 3.991 | 3.98355 | 3.974 | 3.89592 | 3.954 |
| c(Å) | 3.474 | 3.46891 | 3.479 | 3.35961 | 3.431 |
| $B_{iso}(Sr)$ Å$^2$ | 0.47(5) |  | 0.42 |  | 0.44 |
| $B_{iso}(Fe)$ Å$^2$ | 0.47(4) |  | 0.45 |  | 0.45 |
| $B_{iso}(O)$ Å$^2$ | 0.79(5) |  | 0.57 |  | 0.59 |

**Table II.** The calculated zone-centre modes for $SrFeO_2$ and $CaFeO_2$ in meV units. The calculated ab-initio calculations from literature [7] are also given for comparison.

|  | Ab-initio [7] $SrFeO_2$ | Shell model $SrFeO_2$ | Shell model $CaFeO_2$ |
|---|---|---|---|
| $A_{2u}$(TO) | 19.73 | 19.66 | 24.19 |
|  | 34.25 | 37.26 | 37.37 |
| $A_{2u}$(LO) |  | 21.42 | 25.35 |
|  |  | 68.43 | 72.69 |
| $B_{2u}$ |  | 8.20 | 4.82 |
| $E_u$(TO) | 22.71 | 22.43 | 28.24 |
|  | 37.35 | 37.30 | 38.28 |
|  | 65.89 | 66.35 | 67.08 |
| $E_u$(LO) |  | 24.39 | 29.50 |
|  |  | 66.25 | 67.08 |
|  |  | 69.06 | 73.75 |

FIG. 1. Crystal structure of SrFeO$_2$ and CaFeO$_2$. The circles denote Sr/Ca, Fe, and O atoms in decreasing order of size.

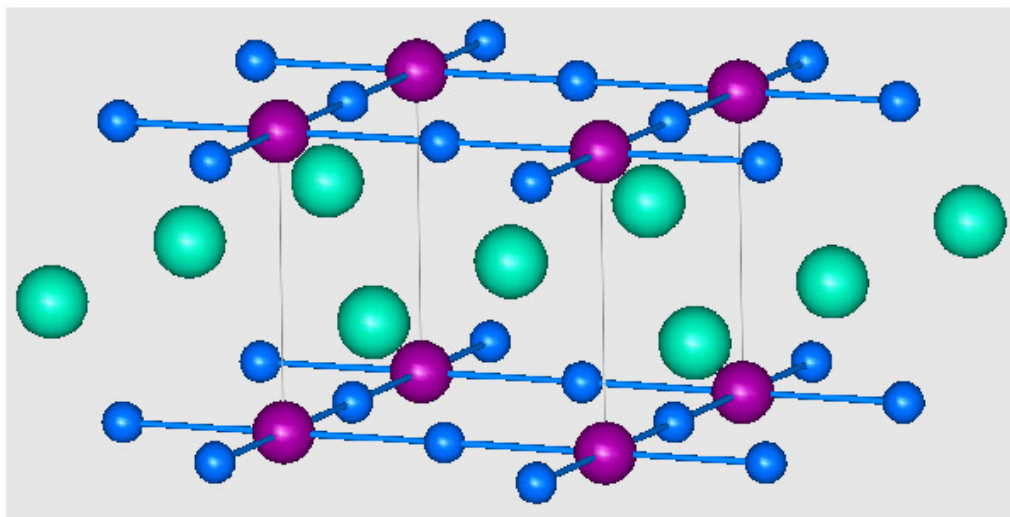

FIG. 2. Calculated phonon dispersion relation in SrFeO$_2$.

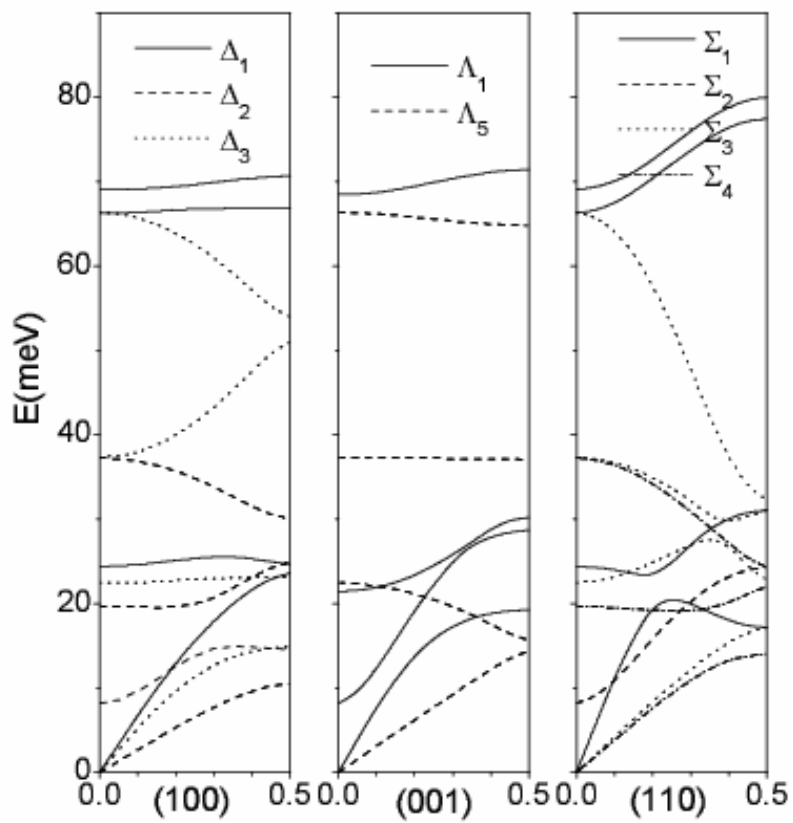

FIG. 3. Calculated partial phonon densities of states of various atoms in $SrFeO_2$ and $CaFeO_2$.

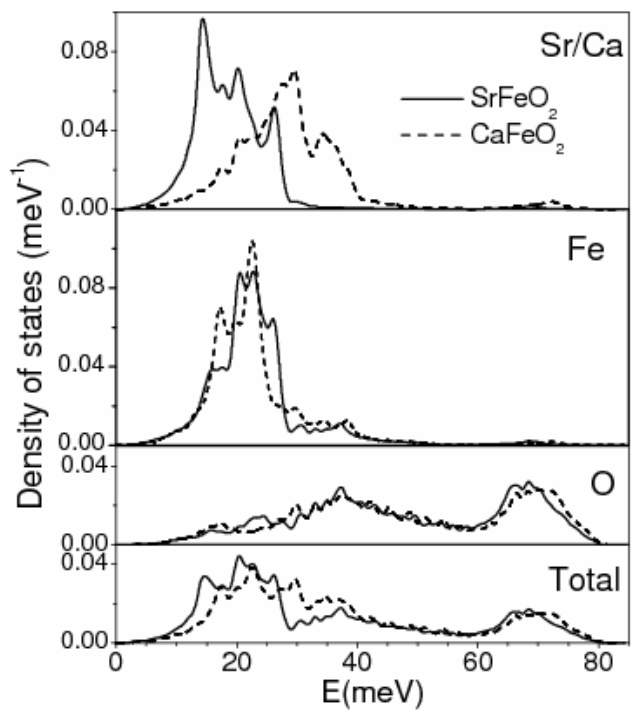

FIG. 4. Calculated equation of state and Debye temperature for $SrFeO_2$ and $CaFeO_2$.

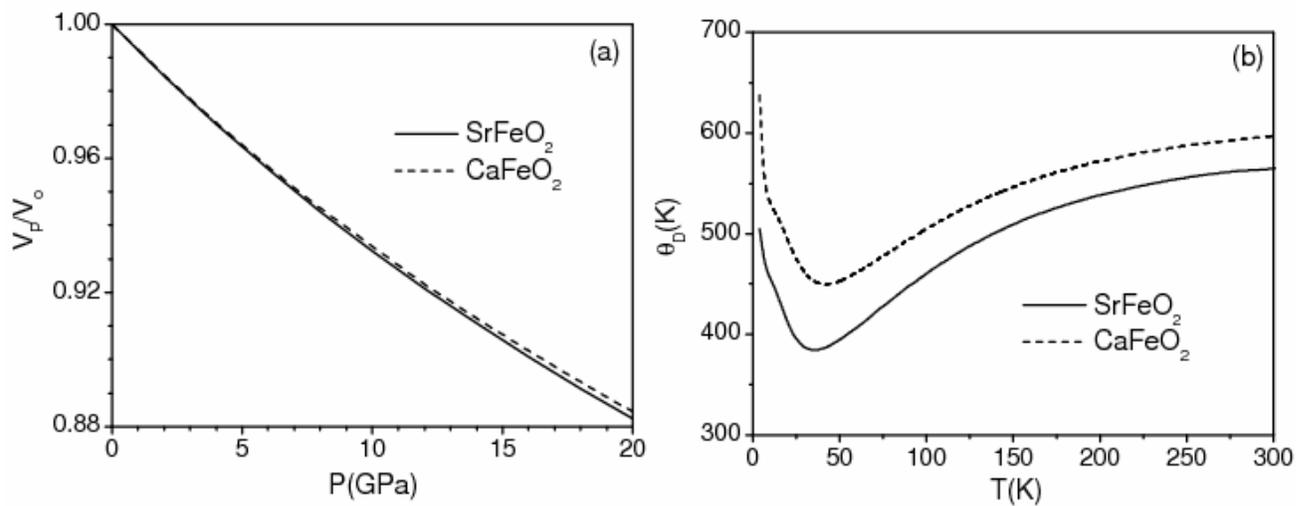

FIG. 5. (a) Calculated Grüneisen parameter $\Gamma(E)$ averaged over phonons of energy E. (b) Comparison between the calculated (lines) and experimental (symbols from Ref. [5]) thermal expansion in $SrFeO_2$ and $CaFeO_2$.

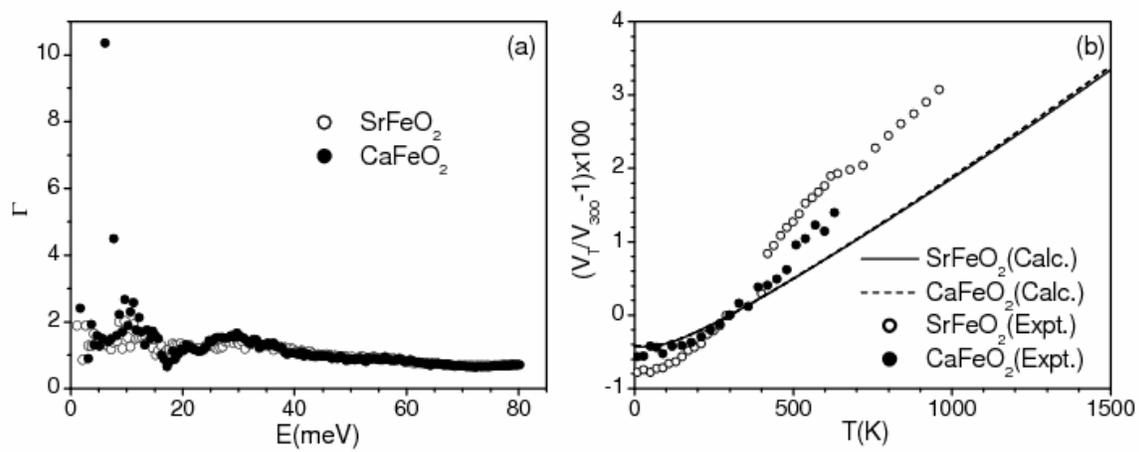